\documentclass{rmaa}

\title{The extreme Type I planetary nebula M2-52\altaffilmark{1}}

\altaffiltext{1}{Based on data collected at the Observatorio Astron\'omico Nacional in San Pedro M\'artir, B.~C., M\'exico.}

\author{Miriam Pe\~na and Selene Medina
  \affil{Instituto de Astronom\'{\i}a, Universidad Nacional Aut\'onoma de M\'exico} } 

\fulladdresses{
\item  M. Pe\~na and S. Medina: Instituto de Astronom\'{\i}a,
  Universidad Nacional Aut\'onoma de M\'exico, 
  Apdo. Postal 70 264, 04510 M\'exico D.F.,
  M\'exico (miriam,selene@astroscu.unam.mx)
  }

\shortauthor{Pe\~na \& Medina}
\shorttitle{The Type I planetary nebula M2-52}

\SetVolume{38} \SetFirstPage{1} \SetYear{2002}

\ReceivedDate{2001 }
 
\AcceptedDate{2002} 

\resumen{ Se presentan los resultados obtenidos a partir de espectroscop{\'\i}a
de alta re\-so\-luci\'on de la parte central de la nebulosa planetaria bipolar
M2-52 que muestra un tipo morfol\'ogico Br.  Hemos confirmado que M2-52 es una nebulosa de
Tipo I de Peimbert, con un espectro rico en l{\'\i}neas de alto y bajo grado de
ionizaci\'on y un fuerte en\-ri\-que\-ci\-mien\-to de He y N.  La composici\'on
qu{\'\i}mica del gas ionizado es:  He/H = 0.165$\pm$0.010, O/H =
(2.6$\pm$0.5)$\times$10$^{-4}$, N/O = 2.3$\pm$0.3, Ne/O = 0.37$\pm$0.10, Ar/O =
(9.2$\pm$2.0)$\times$10$^{-3}$ y S/O~$>$2.0$\times$10$^{-3}$.  La velocidad de
expansi\'on de la nebulosa es, en promedio, de $20\pm2$ km s$^{-1}$ y var{\'\i}a
li\-ge\-ra\-men\-te dependiendo del i\'on considerado.  Los iones de menor grado
de ionizaci\'on, N$^+$ y S$^+$, muestran v$_{\rm exp}$ $\sim$18 km s$^{-1}$,
O$^{++}$ y He$^+$ muestran v$_{\rm exp}$ $\sim$ 20 km s$^{-1}$, en tanto que
He$^{++}$ y H$^+$ muestran v$_{\rm exp}$ $\sim$ 22 km s$^{-1}$.  Es posible que
la zona de N$^+$ y S$^+$ est\'e siendo frenada por el anillo de material
molecular encontrado alrededor de la estrella.}

\abstract{ High-resolution spectrophotometric data of the central zone of the Br-type
planetary nebula M2-52 are presented.  The nebula has a rich spectrum, with high and
low excitation lines.  The chemical composition derived from the spectra shows that He and
N are very enhanced in M2-52.  Thus, this object can be classified as an extreme
Peimbert's Type I PN.  The chemical composition of the ionized gas is:  He/H =
0.165$\pm$0.010, O/H = (2.6$\pm$0.5)$\times$10$^{-4}$, N/O = 2.3$\pm$0.3, Ne/O =
0.37$\pm$0.10, Ar/O = (9.2$\pm$2.0)$\times$10$^{-3}$ and S/O~$>$2.0$\times$10$^{-3}$.  The
expansion velocity of the nebula is, on average, about $20\pm2$ km s$^{-1}$, but the low
ionization species (N$^{+}$ and S$^{+}$) seem to show systematically slightly lower expansion
velocities (18 km s$^{-1}$) than O$^{++}$ and He$^+$ which have v$_{\rm exp}$ = 20 km
s$^{-1}$ while H$^+$ and He$^{++}$ present v$_{\rm exp} \sim 22$ km s$^{-1}$.  This
behavior could indicate that the outer zones of the ionized gas are being decelerated by
the molecular ring located around the central star.  }


\begin{document}

\maketitle

\section{Introduction} \label{sec:intro} 

M2-52 (PN G 103.7+00.4) is a bipolar planetary nebula classified by Manchado et al.  (1996)
as a {\bf Br-type} (bipolar with a ring) nebula.  The nebula shows faint extensions beyond
a central ring.  According to Manchado et al.  (1996) its dimensions are:  total diameter
of 60$''$ and central ring diameter of 23$''$.  Acker et al.  (1992) have reported a flux
at H$\beta$, not corrected for reddening, F(H$\beta$)= 5.01 $\times 10^{-13}$ erg cm$^{-2}$
s$^{-1}$; this flux was computed by assuming a diameter of 14$''$ for the whole nebula.

As other Br-type planetary nebulae (PNe) studied to the present, M2-52 has been found to
have an important amount of molecular material in the ring.  Guerrero et al.  (2000) have
found a large amount of H$_2$ with a total flux of F(H$_2$)= 1.98$\times 10^{-12}$ erg
cm$^{-2}$ s$^{-1}$.  The same authors established that, in comparison with the Br$\gamma$
emission, M2-52 appears as a {\it H$_2$-dominated PN}, with a ratio F(H$_2$)/F(Br$\gamma$)
= 8.6.  In addition Zhang et al.  (2000) have reported the detection of molecular CO(1--0)
in this object, with an intensity corresponding to a molecular mass of about 0.085
M$_\odot$ if they assume a distance of 4.2 kpc.  Guerrero et al.  (2000) and Zhang et al.
(2000) found that the molecular material is located in the ring and concentrated in two
bright knots separated (from peak to peak) by about 7$''$.  Zhang et al.  derived a radial
velocity of $-$63.2 km s$^{-1}$ for this object.

 
An expansion velocity of 7.5 km s$^{-1}$ has been measured by Sabbadin et al.  (1985), by
fitting two gaussian distributions to the [\ion{O}{3}] 5007 line which appears single and
broad.


Condon \& Kaplan (1998) presented radio observations at 1.4 GHz of a great number of
galactic planetary nebulae.  For M2-52 they have reported a flux $S_{\nu}$ = 15.4$\pm$0.6 mJy,
which they combined with the total flux at H$\beta$ to compute the logarithmic reddening
correction at H$\beta$, obtaining a value of c(H$\beta$)= 1.0.

From optical spectroscopy, Kaler et al.  (1996) found that M2-52 can be classified as
a Peimbert's Type I planetary nebula.  Type I PNe are characterized for showing large
He- and N-enrichment and presumably they evolve from the most massive PN progenitor
stars.  These objects are potentially an important source of He and N enrichment in the
interstellar medium.  A substantial fraction of Type I PNe shows bipolar morphology
and a notorious filamentary structure (Peimbert 1978; Peimbert 1985 and references
therein). 

In this work we present high resolution spectrophotometric data for the central zone of M2-52
demonstrating that it is an extreme Type I planetary nebula, comparable to the
outstanding NGC\,2440 and NGC\,2818.  In section 2, we describe the observations
and data reduction.  The analysis of kinematics and photometric data is
discussed in Section 3, and in Section 4 we present our results.

\begin{table*}[th!]
\caption{Observed and dereddened fluxes of M2-52, relative to H$\beta$}
\begin{flushleft}
\begin{center}
\begin{tabular}{lcrrrrrrr}
\hline
\noalign{\smallskip}
  &  &  &\multicolumn{2}{c}{center zone}&\multicolumn{2}{c}{east knot}& \multicolumn{2}{c}{all}\\
  &  &  &\multicolumn{2}{c}{($4''\times 3''$)}&\multicolumn{2}{c}{($4''\times 3''$)}&\multicolumn{2}{c}{($4''\times 13''$)}\\ 
ion& $\lambda$ & f$_\lambda$~~ & F$_\lambda$/F(H$\beta$) & I$_\lambda$/I(H$\beta$) & F$_\lambda$/F(H$\beta$) & I$_\lambda$/I(H$\beta$) & F$_\lambda$/F(H$\beta$)&I$_\lambda$/I(H$\beta$)\\
\hline 
{[\ion{O}{2}]}& 3726 &   0.256 & 0.42:  &  0.85: & 0.56& 1.36 & 0.50  & 1.14\\
{[\ion{O}{2}]}& 3729 &   0.255 & 0.40: &  0.81: & 0.54& 1.30 & 0.55: & 1.25:\\
{[\ion{Ne}{3}]}& 3869 &   0.223 & 0.90 & 1.67 & 1.02  & 2.16 & 0.93 & 1.91 \\
{[\ion{Ne}{3}]}& 3967 &   0.203 & 0.32 & 0.65  & --- & --- & --- & --- \\
{[\ion{S}{2}]} & 4069 &   0.178 & ---  & --- & --- & --- & 0.06: & 0.11:\\
\ion{C}{2}    & 4267 &  0.141 & --- & --- & --- & --- & $<$0.01 & $<$0.02 \\
H$_\gamma$   & 4340  &  0.125& 0.34 &  0.47 & 0.30 & 0.46 & 0.32 & 0.46\\
{[\ion{O}{3}]}& 4363 &   0.124 & 0.13 &  0.18  & 0.15& 0.24 & 0.15 & 0.22\\                                                                                                                                              
\ion{He}{2}  & 4686  &   0.042 & 0.83 &  0.93 & 0.75& 0.87 & 0.72  & 0.83 \\
{[\ion{Ar}{4}]}& 4711 &   0.039 & 0.11: & 0.12: & 0.12 & 0.13& 0.09 & 0.10\\
{[\ion{Ne}{4}]}& 4725 &   0.035 & 0.02:& 0.02: & --- & --- & 0.01: & 0.02: \\
{[\ion{Ar}{4}]}& 4741 &   0.031 & 0.09: & 0.10: & 0.10 & 0.11& 0.08 & 0.08 \\
{[\ion{O}{3}]}& 4959 &   0.023 & 4.30  &  4.03  & 4.50 & 4.16  & 4.04  & 3.75 \\
{[\ion{O}{3}]}& 5007 & $-0.033$& 12.90  & 11.77 & 14.10 & 12.58  & 13.5 & 12.13 \\
{[\ion{N}{1}]}& 5200 & $-0.073$& 0.11: &  0.09: & --- & ---& 0.16: & 0.12:\\
\ion{He}{2}  & 5411  & $-0.118$& 0.12 &  0.08 & 0.12 & 0.09 & 0.11 & 0.07\\
{[\ion{N}{2}]} & 5755 & $-0.185$& 0.22 & 0.13 & 0.40 & 0.21 & 0.34 & 0.19\\
\ion{He}{1}  & 5876  & $-0.208$& 0.18 &  0.10  & 0.25& 0.12 & 0.23 & 0.12\\
{[\ion{O}{1}]}& 6300 & $-0.284$& 0.14 &  0.06 & 0.25& 0.09 & 0.36 & 0.14\\
{[\ion{S}{3}]} & 6312 & $-0.286$& 0.16 & 0.07 & 0.21 & 0.08 & 0.18 & 0.06\\
H$_\alpha$   & 6563  &  $-0.330$ & 7.10  &  2.83  & 8.75 & 2.80 & 8.14 & 2.81\\  
{[\ion{N}{2}]} & 6583 & $-0.335$& 15.14 & 6.00  & 27.92 & 8.78 & 23.90 & 8.13\\
{[\ion{S}{2}]} & 6717 & $-0.343$& 0.92 & 0.36 & 1.62 & 0.50& 1.61 & 0.53\\
{[\ion{S}{2}]} & 6731 & $-0.344$& 0.92 & 0.35 & 1.69 & 0.52 & 1.61 & 0.53\\
{[\ion{Ar}{5}]}& 7006 & $-0.375$& 0.19  & 0.07  & 0.14 & 0.05 & 0.14& 0.04\\
{[\ion{Ar}{3}]}& 7136 & $-0.390$& 1.07  & 0.36 & 1.17 & 0.26 & 1.15 & 0.37\\
\hline
c(H$\beta$) &  & & 1.2$\pm$0.2 &  & 1.5$\pm$ 0.2 & &1.4$\pm$0.2 \\
\multicolumn{3}{l}{log F(H$\beta$) (erg cm$^{-2}$ s$^{-1}$})  & $-14.00$ &  & $-13.82$ & & $-13.25$\\
\hline
\end{tabular}
\end{center}
\end{flushleft}
\end{table*}

\begin{table*}[t]
\caption{Physical conditions and chemical composition}
\begin{center}
\begin{tabular}{lrrr}
\hline
 & center zone &east knot& all~~~~~\\
\hline
T[\ion{O}{3}]    &         14800$\pm$1200 &   14600$\pm$1100 &   14600$\pm$1000 \\
T[\ion{N}{2}]     &         12300$\pm$1200 &   12800$\pm$1200 &   12500$\pm$1000 \\
N[\ion{S}{2}]     &           600$\pm$400 &     800$\pm$400 &     700$\pm$400  \\
N[\ion{A}{4}]     &          1700$\pm$800 &    1200$\pm$800 &     800$\pm$600  \\
N[\ion{O}{2}]      &         610: &    700: &    500:  \\
He$^+$ (5876)  &    7.20E-02  &8.55E-02 & 8.40E-02 \\
He$^{++}$ (4686) &  9.10E-02  &8.49E-02 & 8.06E-02 \\
N$^+$ (6583)   &    5.87E-05  &8.26E-05 & 7.73E-05 \\
O$^+$ (3727)   &    2.47E-05  &3.69E-05 & 3.34E-05 \\
O$^{++}$ (5007) &   1.27E-04  &1.42E-04 & 1.36E-04 \\
Ne$^{++}$ (3869)&   4.24E-05  &5.78E-05 & 5.05E-05 \\
Ne$^{+3}$ (4725)&   2.56E-05  &2.79E-05 & 2.19E-05 \\
S$^+$    (6730) &   8.59E-07  &1.23E-06 & 1.27E-06 \\
S$^{++}$  (6312)&   4.16E-06  &4.70E-06 & 3.85E-06 \\
Ar$^{++}$  (7136)&   1.48E-06  &1.09E-06 & 1.54E-06 \\
Ar$^{+3}$  (4740)&   6.68E-07  &7.33E-07 & 5.47E-07 \\
Ar$^{+4}$  (7006)&   4.72E-07  &3.29E-07 & 2.72E-07 \\
He/H           &    0.163$\pm$0.010  &0. 170$\pm$0 010 & 0.165$\pm$0.010 \\
O/H            &    (2.6$\pm$0.5)E-04  &(2.8$\pm$0.5)E-04 & (2.6$\pm$0.5)E-04 \\
N/O (N$^+$/O$^+$)&  2.4$\pm$0.3  &2.2$\pm$0.3 & 2.3$\pm$0.3 \\
Ne/O (Ne$^{++}$/O$^{++}$)&0.33$\pm$0.15 &0.41$\pm$0.15 & 0.37$\pm$0.15 \\
S/H (S$^+$+S$^{++}$) & $>$5.0E-06 & $>$5.9E-06 & $>$5.1E-06\\
Ar/H (Ar$^{++}$+Ar$^{+3}$+Ar$^{+4}$)& (2.7$\pm$0.5)E-06 & (2.2$\pm$0.4)E-06 & (2.4$\pm$0.5)E-06 \\
\hline
\end{tabular} 
\end{center}
\end{table*}

 \section{Observations and data reduction} 

Two high-resolution {\it echelle} spectra with exposure times of 10 min and 20
min, respectively, were obtained on 2000 November 2, with the 2.1-m telescope
and the Thomson TH7398M CCD (2048$\times$2048 pixels of
14$\mu$m$\times$14$\mu$m), at the Observatorio Astron\'omico Nacional (OAN), San
Pedro M\'artir, B.C., M\'exico.  Slit dimensions were 4$''$  along the dispersion
and 13.3$''$ along the spatial coordinate. The slit was
E-W oriented, and the spectral range covered was from 3300 to 7300 \AA, with a
spectral resolution between 0.1 and  0.2 \AA.  Th-Ar comparison lamps were used for
wavelength calibration and three standard stars from the list of Hamuy et al.
(1992) were observed for flux calibration.

Due to the large extension of M2-52, our slit included only the central zone and part of the
bright ring around the central star.  The bipolar nature of M2-52 is evident in our
bi-dimensional spectra.  Figure 1 shows the structure of the [\ion{N}{2}] 6583 emission
line.  The emission is extended, filling the slit along the spatial axis, and two bright
knots are clearly detected, separated by 6.5 arcsec.  The stellar continuum emission should
be located in the middle, between the knots, but the star is too faint to be detected in our
high resolution  spectrum.  All the nebular lines present in our spectrum show
a similar structure, except \ion{He}{2} 4686 for which the knots appear less defined. That is, the \ion{He}{2} 
emission is more concentrated towards the central zone nearer the central star. The
knotty structure we have detected agrees well with the morphology presented by Guerrero et al.  (2000, see
their Fig.  1b).  The knots observed in [\ion{N}{2}] almost coincide with the bright knots
in H$_2$ found by them.  Both emissions arise from the ring of the nebula.

\begin{figure}[t]
\begin{center}
\includegraphics[height=8cm,width=7cm] {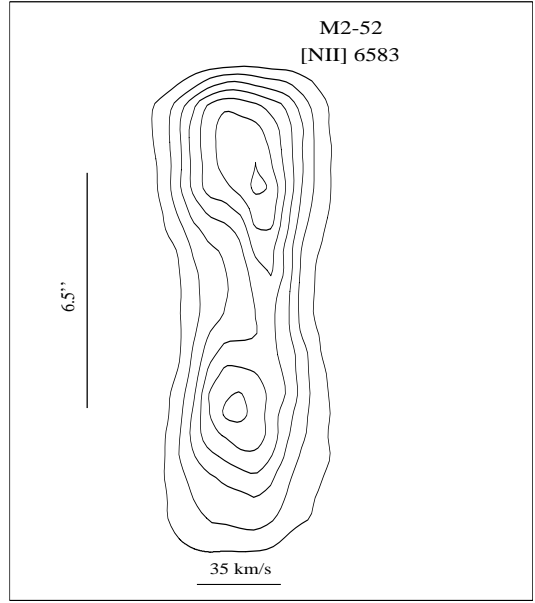}

\caption{Contour diagram of the [\ion{N}{2}] 6583 emission line in the bi-dimensional
spectrum.  The spatial direction is along the y axis.  The two knots detected are oriented
E-W (East is up), and separated by 6.5$''$.  The stellar continuum should be located
between the knots, but the star is too faint to be detected in our {\it echelle} spectrum.
The emission is filling the slit of 13.3$''$ length.  The FWHM of the [\ion{N}{2}] line is
0.95 \AA, equivalent to a velocity range of 35 km s$^{-1}$.}  
\label{fig1} 
\end{center}
\end{figure}

Our bi-dimensional {\it echelle} spectra were bias-subtracted and flat-fielded
using IRAF standard procedures\footnote{IRAF is distributed by NOAO, which is
operated by AURA, Inc., under contract with the NSF.}.  Then we have proceeded to
extract spectral data of three different zones:  The central zone between the knots was extracted
with an aperture of $4'' \times 3''$ (it would correspond to the zone nearest to
the central star); the  emission from the East knot was extracted with an aperture of also $4''
\times 3''$ and, finally we extracted almost all the nebular emission in our slit, with an aperture of $4''
\times 13''$. Extracted spectra were 
 wavelength and flux calibrated.  Our final
spectra are an average of the two observations.

In Table 1 we present the observed fluxes, F$_\lambda$, and the dereddened fluxes,
I$_\lambda$, relative to H$\beta$, for the most important lines detected in the three
zones.  The dereddened fluxes were derived from the observed fluxes employing a logarithmic
reddening correction at H$\beta$, $c(\rm H\beta$), as derived for each zone, from the
Balmer decrement by considering case B recombination theory (Hummer \& Storey 1987).  We
used the reddening law given by Seaton (1979), which is listed in column 3 of Table 1
(f$_\lambda$).  The values for $c(\rm H\beta$) are given at the bottom of Table 1.  In the
three regions we found similar reddening coefficients (within uncertainties), though it is
interesting to notice that the knot presents a slightly larger reddening, probably due to
the molecular material located near this region.

As expected from the ionization structure of photoionized nebulae, the line intensities 
of low ionization species appear larger in the knot, far from the central star, than in the
central zone.

Uncertainties in line ratios were determined by comparing the measurements of our two
spectra.  In a general way, the uncertainties for lines with F$_\lambda$/F(H$\beta$)
$\geq$ 0.2 are better than 10\% and improving with the line flux (for instance, the
uncertainties for [\ion{O}{3}] 5007 are about 3\%).  The exceptions are [\ion{O}{2}] 3726
and 3729 for which uncertainties of about 20\% are found.  This is because the large
reddening affecting M2-52 weakens the UV lines.  Lines with 0.2 $>$
F$_\lambda$/F(H$\beta$) $>$ 0.05 (such as the important [\ion{O}{3}] 4363 and [\ion{N}{2}]
5755), have uncertainties of about 20\% and the uncertainties are larger for lines marked
with a colon.  The fluxes at H$\beta$, as measured with the different extraction
apertures, are given at the end of Table 1.

Plasma diagnostics were performed from the emission line ratios in
a standard way, using the same atomic data as listed in Stasi\'nska \&
Leitherer (1996).

Electron densities were derived from [\ion{O}{2}] 3726/3729, [\ion{S}{2}]
6717/6731 and [\ion{Ar}{4}] 4711/4740 line ratios, electron temperatures were measured from
[\ion{O}{3}] 4363/5007 and [\ion{N}{2}] 5755/6583 ratios.  The
density used for deriving the electron temperatures was that deduced
from [\ion{S}{2}] 6717/6731, which is more confident and equal within
uncertainties to the density obtained from the [\ion{Ar}{4}] and [\ion{O}{2}] lines.

The derived electron temperatures and densities are listed in Table 2, together with the
errors based on the uncertainties of the line ratios described above.  We do not find
any  systematic difference in the temperatures of the different regions.
Temperatures and densities are equal in the central zone and the knot within uncertainties,
although the knot could be slightly denser.

Ionic abundances were then obtained, for the three regions, using T[\ion{O}{3}] for the
high ionization species and T[\ion{N}{2}] for the low ionization ones.  Electron densities
derived from [\ion{S}{2}] 6717/6731 ratios were always used.  The results are presented in
Table 2, where we also indicate which emission line has been employed to derive the ionic
abundance.  No temperature fluctuations were considered in deriving the ionic abundances,
therefore, they should be considered as lower limits of the true chemical abundances (See
Peimbert et al.  1995 for a discussion of the effects of temperature fluctuations on
chemical abundance determinations).  However, due to the low density in the nebula, we do
not expect very large deviations of the derived chemical abundances.

Elemental abundance ratios were computed from the ionic abundance ratios using
the ionization correction factors of Kingsburgh \& Barlow (1994).  The abundance
ratios He/H, O/H, N/O, Ne/O, S/H and Ar/H are given at the bottom of Table 2.
The value for S/H is the addition of (S$^+$ + S$^{++}$)/H$^+$, while the value
for Ar/H is the addition of Ar$^{++}$, Ar$^{+3}$ and Ar$^{+4}$ abundances,
therefore only Ar/H can be considered a confident value.  The error bars that
are listed for O, N, Ne and Ar abundances take into account the uncertainties
propagated from the uncertainties in the physical conditions (electron
temperature and density).

\section{Data analysis and discussion} 

\subsection{Kinematics}

Our {\it echelle} spectra have a resolution better than 10 km s$^{-1}$ in
average, which is good enough to derive the radial and expansion velocities of the
nebula.  We find that the heliocentric radial velocity, as measured from all the available
lines, is $-73\pm7$ km s$^{-1}$, in good agreement with the values presented
in previous works.

Analyzing the structure of [\ion{N}{2}] 6583 line shown in Fig.  1, it is found that the
knots, situated at each side of the central star, do not show a significant difference in
velocity (they appear well aligned along the y axis) and the same is found for the
other ions.  If the knots belong to the ring which is forming the ``waist'' of the bipolar
structure, as suggested by Guerrero et al.  (2000), such a ring does not present important
expansion nor rotation.

We have analyzed the line profiles of the ions present in the gas for the central zone
spectra.  All the lines appear single (not split) but well resolved and we have measured
the FWHM of each line, proceeding then to subtract the instrumental width.

It is usual in the literature to interpret the separation of double-peak lines or the FWHM
of single lines as expansion velocity of the nebular shell, although a certain amount of
turbulence and other parameters like density and thermal structures, could also be
contributing to the line shapes and widths.  Gesicki et al.  (1996) and Neiner et al.
(2000) have demonstrated that for the case of PNe ionized by non-[WC] central stars (as the
case of M2-52) the turbulent velocity field is negligible and the expansion velocity
increases from high to low ionization species.  

We have  determined the expansion
velocities of the different ions present in M2-52 from the FWHM of their lines.  These
values are listed in Table 3.  The uncertainties in this table have been computed by taking
into account the measurements, in both spectra, of the available emission lines for each
ion.

\begin{table}
\caption{Expansion velocities}
\begin{center}
\begin{tabular}{lc}
\hline
ion & 2v$_{\rm exp}$ (km s$^{-1}$) \\
\hline
\ion{He}{2} 4686,5411 & 45$\pm$4 \\
\ion{He}{1} 5876 & 38$\pm$7 \\
{[\ion{O}{3}]} 5007,4959 & 39$\pm$4 \\
\ion{H}{1} H$\alpha$,H$\beta$,H$\gamma$ & 45$\pm$4 \\
{[\ion{N}{2}]} 5755,6548,6583 & 36$\pm$3 \\
{[\ion{S}{2}]} 6717,6731 & 36$\pm$4 \\
\hline
\end{tabular}
\end{center}
\end{table}

Within uncertainties all the ions present similar expansion velocities, and an average
v$_{\rm exp} \sim 20\pm2$ km s$^{-1}$ could be adopted for the nebula.  Nevertheless, it is
notorious that low ionization species such as N$^{+}$ and S$^{+}$ show systematically lower
expansion velocities ($18\pm3$ km s$^{-1}$) than He$^{++}$ and H$^+$, for which v$_{\rm
exp}$ $\sim$ 22$\pm$4 km s$^{-1}$.  Also O$^{++}$ and He$^+$ show lower expansion
than He$^{++}$ and H$^+$. This behavior is opposite to the one found by Gesicki et al.
 (1996) and Neiner et al. (2000) and should 
 be verified with better
resolution spectroscopic data.  If real, it could be indicating that the 
less ionized zones of the nebula (where N$^{+}$ and S$^{+}$ are located) are being
decelerated by the torus of molecular material around the central star.  It would be
interesting to verify if such a behavior is found in other Br-type PNe.

Our value of v$_{\rm exp}$ for M2-52 is larger than the one reported by Sabbadin et al.
(1985).  This is due to the different methods employed for measuring.  We have fit
single gaussians to the line profiles because, as we said before, the lines appear single
and one gaussian distribution is an adequate fit.  Sabbadin et al.  (1985) have adjusted two gaussians to
the single profile, taking 2v$_{\rm exp}$ as equal to the difference in velocity of the two
maxima of both gaussians.  Their procedure produces a lower v$_{\rm exp}$.

\subsection{Ionic and total abundances}

M2-52 shows an emission line pattern similar to those of other Type I PNe, in the sense
that high and low ionization lines are present.  We have detected lines of Ar$^{+3}$ as
well as of O$^0$ in the central zone and also in the knot, although the knot presents a
lower excitation and the lines of low ionization species are enhanced in this zone.  This
difference in excitation is due to the normal ionization structure of a nebula
photoionized by a hot central star.  The highly ionized gas is closer to the  star.

The ionization degree in M2-52 is very high, as it is deduced by the large
fraction of He twice ionized (He$^{++}$/He = 0.50 in the knot and 0.56 in the center).
This indicates a high effective temperature for the central star (certainly hotter than
80,000 K).  Unfortunately the central star is faint and in our high-resolution spectra we
have not detected the stellar emission, thus no Zanstra temperature could be derived.

The elemental abundances of M2-52, presented in Table 2, are equal within uncertainties in 
the central zone and in the knot and, in the following, we will adopt the values obtained
for the whole nebula for our analysis.  As shown in Table 2, M2-52 is a very He- and N-rich
nebula.  The high He/H ratio of 0.165 is one of the largest reported for Type I PNe and it
is only comparable to the values of NGC\,2818 (Peimbert \& Torres-Peimbert 1987) and
He\,2-111, the extreme Type I PN reported by Kingsburgh \& Barlow (1994).  Also the N/O
ratio of 2.3 is one of the largest reported, and it is similar to the large N/O value of
NGC\,2440, considered as the prototype of Peimbert's Type I PNe.  For the latter nebula,
Hyung \& Aller (1998) have computed a N/O ratio of 2.16, which they consider that could be
an artifact produced by an ``abnormally'' strengthened [\ion{N}{2}] lines in the blobs of
NGC\,2440.  Other similarities between M2-52, NGC\,2818 and NGC\,2440 are notorious, for
instance the electron temperatures are very similar in these objects (Peimbert et al.
1995), indicating similar heating and cooling processes.  Also the morphology of M2-52 is
very similar to that of NGC\,2818 and NGC\,2440.

Comparing our values for M2-52 with those by Kaler et al.  (1996), we found that our line
ratios and physical conditions are similar to theirs, except for the logarithmic reddening
correction c(H$\beta$) for which they derive a value of 1.6, larger than our $1.2\pm0.2$
for the center and $1.4\pm0.2$ for the whole nebula.  This discrepancy made that Kaler et
al. have found a lower N/O ratio, due to the larger extinction correction applied to their
[\ion{O}{2}] 3727 doublet.  A value for c(H$\beta$) can be estimated from the observed
intensity of Br$\gamma$ ($2.3\times10^{-13}$ erg cm$^{-2}$ s$^{-1}$) given by Guerrero et
al.  (2000), relative to the total H$\beta$ flux given by Acker et al.  (1992).  By
adopting a theoretical Br$\gamma$/H$\beta$ ratio of 0.028 (Osterbrock 1989) and a reddening
correction factor for Br$\gamma$, f$_\lambda = -0.901$ (Cardelli et al.  1989) we have
deduced c(H$\beta$)= 1.36, in very good agreement with our value from the Balmer decrement.
Also our value of c(H$\beta$) is similar to the reddening reported by Condon \& Kaplan
(1998) from radio observations of M2-52.  Therefore we are confident in our results.

As in other Type I PNe, the extreme He- and N-enrichment in M2-52 indicate that the central
star experienced envelope-burning conversion to nitrogen of primary carbon extracted during
the third dredge-up event.  Unfortunately we were not able to determine the carbon abundance of
M2-52, due to the weakness of the \ion{C}{2} 4267 emission line, for which only an upper
limit is given.  It is possible that carbon is also enhanced in M2-52, as it is in
NGC\,2440 (Hyung \& Aller 1998), due to the notorious similarities between physical
conditions and chemical composition in both nebulae.

\section{Conclusions} 

The Br-type nebula M2-52 appears to be a high excitation Type I PN with a rich spectrum
including high and low excitation lines.  From high resolution spectrophotometry of the
central zone, we have found that the ionized shell shows an expansion velocity of about $20\pm2$
km s$^{-1}$.  The low ionization species, however, seem to have slightly lower expansion velocities
than the high ionization ones.  This could indicate that the ring of molecular material,
detected around the central star, is decelerating the expansion of the external zones of
the ionized gas.  Better resolution spectroscopy is required to confirm this result.

The physical conditions in the ionized gas are:  T[\ion{O}{3}] = $14\,600\pm1\,000$ K,
T[\ion{N}{2}] = $12\,500\pm1\,000$ K, electron density $\sim$ 800
cm$^{-3}$, and the chemical composition is:  He/H =
0.165$\pm$0.010, O/H = (2.6$\pm$0.5)$\times 10^{-4}$, N/O = 2.3$\pm$0.3, Ne/O =
0.37$\pm$0.15, Ar/O = (9.2$\pm$2.0)$\times$10$^{-3}$, and S/O $>$ 2.0$\times$10$^{-3}$.

Our oxygen abundance for M2-52 is a relatively low value when compared with the O/H
abundances derived for other Type I PNe by Peimbert et al.  (1995).  This is understandable
when we consider that Peimbert et al.  values include the effects of temperature
fluctuations.  On the other hand, the abundance ratios relative to oxygen are not as
affected by this phenomenon because the ratios are not largely temperature dependent,
therefore the N/O, Ne/O and Ar/O ratios are confident values.  The large He/H and N/O
ratios clearly establish that M2-52 is a Peimbert's Type I PN presenting one of the largest
He and N enrichments, similar to the values for NGC\,2440 and NGC\,2818.  Neon and argon
abundances in M2-52, relative to oxygen, are in very good agreement with the bulk of Type I
PNe (Kingsburgh \& Barlow 1994; Peimbert et al.  1995).

\acknowledgments

This work received partial support from DGAPA/UNAM (grant IN100799) and CONACYT (grant
35594-E).  S.M.  acknowledges scholarship from DGEP/UNAM and CONACYT.  


\end{document}